\documentclass[10pt,journal]{IEEEtran}

\usepackage[scaled=0.86]{berasans}  
\usepackage[colorlinks=true, allcolors=blue, urlcolor=blue]{hyperref}  
\usepackage{graphicx} 
\usepackage[babel]{microtype}  
\usepackage{amsmath,amssymb,amsthm,bm,amsfonts,mathrsfs,bbm} 

\usepackage{xspace}  
\usepackage{pgfplots}
\usepackage{xcolor,colortbl}
\usepackage{array}
\usepackage{bigstrut}
\usepackage{tcolorbox}
\usepackage{algorithm2e}
\usepackage{orcidlink}

\newcommand{\tr}{\text{tr}}

\newcommand{\be}{\begin{equation}}
\newcommand{\ee}{\end{equation}}
\newcommand{\bea}{\begin{eqnarray}}
\newcommand{\eea}{\end{eqnarray}}
\newcommand{\bes}{\begin{equation*}}
\newcommand{\ees}{\end{equation*}}
\newcommand{\beas}{\begin{eqnarray*}}
\newcommand{\eeas}{\end{eqnarray*}}

\newcommand{\ket}[1]{|#1\rangle}
\newcommand{\bra}[1]{\langle#1|}
\newcommand{\proj}[1]{\ket{#1}\!\bra{#1}}

\def\a{\vec{a}}

\def\tr{\mathrm{tr}}

\newtheorem*{thm*}{Theorem}

\newtheorem*{lem*}{Lemma}

\newtheorem*{lipschitzLem*}{Lemma \ref{lipschitz}}
\newtheorem*{lipschitzCubeLem*}{Lemma \ref{lipschitzCube}}
\newtheorem*{pgmNearlyOptimalThm*}{Theorem \ref{pgmNearlyOptimal}}

\begin{document}
\RestyleAlgo{ruled}

\title{Mitigation of Crosstalk Errors in Quantum Measurements}
 
\author{Seungchan Seo{\orcidlink{0000-0002-2256-8989}}, Jiheon Seong{\orcidlink{0009-0002-5225-8305}}, Huan-Yu Ku{\orcidlink{0000-0003-1909-6703}}, and Joonwoo Bae{\orcidlink{0000-0002-2345-1619}}, {\it Member, IEEE}
\thanks{Seungchan Seo, Jiheon Seong, and Joonwoo Bae are with School of Electrical Engineering, Korea Advanced Institute of Science and Technology (KAIST), 291 Daehak-ro, Yuseong-gu, Daejeon 34141, Republic of Korea (e-mail: ichi9505@kaist.ac.kr; jiheon94@kaist.ac.kr; joonwoo.bae@kaist.ac.kr)}
\thanks{Huan-Yu Ku is with the Department of Physics, National Taiwan Normal University, Taipei, Taiwan. (e-mail: huan.yu@ntnu.edu.tw)}
}






\maketitle

\begin{abstract}


In the present-day quantum information technologies, measurements are performed not only at the end of quantum state evolution but also between quantum circuits to connect short-depth quantum circuits, for instance, variational quantum algorithms. Furthermore, measurements also contain crosstalk errors, which cannot be factorized into individual readout errors. In this work, we investigate methods for mitigating measurement errors across multiple qubits, particularly measurement crosstalk that cannot be corrected by methods for correcting individual readout errors. We present mitigation protocols that generally suppress these errors by up to a percentage level across multiple qubits. We first demonstrate that mitigating measurement errors of individual qubits cannot eliminate measurement crosstalk errors. We then present methods for mitigating measurement readout errors across multiple qubits, accounting for crosstalk noise. Both methods rely on quantum detector tomography. We also present proof-of-principle demonstrations of measurement error mitigation on IBM quantum hardware. 
\end{abstract}


\section{Introduction}



\IEEEPARstart{Q}{uantum} measurements are a fundamental building block in quantum information processing. They are applied for various purposes: measuring errors, called syndrome measurements for quantum error correction, or reading out results of quantum algorithms \cite{PhysRevA.52.R2493, gottesman}. The currently available quantum technologies also exploit measurements to connect short-depth circuits, such as variational quantum eigensolver or quantum approximate optimization algorithm~\cite{BLEKOS20241, bharti2021noisy, TILLY20221}. All of these find it crucial to tackle noise appearing in quantum measurements. 

So far, methods for mitigating single-qubit measurements have been presented and are based on quantum detector tomography (QDT) or quantum twirling, see e.g., \cite{Maciejewski2020mitigationofreadout, 9142431, Wang:2023aa, Geller_2020, PhysRevLett.127.090502}. It has also been pointed out that for noisy quantum hardware, correlated errors in quantum measurements, known as measurement crosstalk, may be present. Note that measurement crosstalk can be characterized as a noisy measurement on multiple qubits where the measurement cannot be factorized into individual measurements \cite{PhysRevA.100.052315, PhysRevApplied.17.014024, PRXQuantum.2.040338,Sarovar2020detectingcrosstalk}. 

In this work, we present quantum measurement-error mitigation protocols that generally work on suppressing readout errors on multiple qubits. The framework for mitigating measurement errors requires QDT in advance, based on which two stages are devised: one called quantum preprocessing that applies local unitaries before measurements, and the other a classical postprocessing to manipulate raw probabilities of noisy outcomes. We present pre- and post-processings by dealing with noisy measurements in two ways. Firstly, we exploit an eigendecomposition of a noisy measurement and construct a mitigation protocol to wash out the effect of noise. Secondly, we introduce the preferred basis decomposition for a noisy measurement by pre-determining the basis for noiseless measurements. Based on the decomposition, we present the mitigation protocol which suppresses a multi-qubit readout error rate up to $O(10^{-2})$. We apply the proposed mitigation protocols to IBM quantum devices and demonstrate suppressions of measurement errors. 

This work is structured as follows. In Sec. \ref{sec:m}, we summarize noisy quantum hardware in the present-day technologies. Measurement noise is introduced, including crosstalk errors. In Sec. \ref{sec:p}, we present protocols for mitigating measurement noises. We establish the framework of mitigating measurement errors by quantum preprocessing and classical postprocessing. Two protocols are presented and compared. In Sec. \ref{sec:l}, we show that measurement crosstalk errors cannot be mitigated by repeating a protocol designed to mitigate single-qubit measurement errors. In Sec. \ref{sec:a}, we apply mitigation protocols to realistic scenarios via cloud-based quantum computing services and the device `ibm\_brisbanethe fake backend `FakeBrisbane' that mimics real QPU `ibm\_brisbane'. Proof-of-principle demonstrations of measurement error mitigation are presented. In Sec. \ref{sec:c}, we conclude the results and discuss future directions.

\section{ Measurements in noisy quantum hardware }
\label{sec:m}

\subsection{Preliminaries}

In this section, we collect notations and terminologies to be used throughout. In a noiseless scenario, a measurement on a single-qubit state is performed in the computational basis, in which measurement operators are given as
\bea
\mathrm{noiseless~measurement:~ }\{|0\rangle\langle 0|, |1\rangle \langle 1| \}. \label{eq:nno1}
\eea
A general measurement that can describe noisy cases is denoted by a positive-operator-valued-measure (POVM). Let $\Pi_0$ and $\Pi_1$ denote POVM elements of a noisy measurement giving outcomes $0$ or $1$,
\bea
\mathrm{ noisy~ measurement:~ } \{\Pi_0, \Pi_1 \}. \label{eq:no1} 
\eea
Note also that, throughout, a measurement is complete; hence, we only have outcomes $0$ or $1$. 
Then, a POVM element of a noiseless measurements on multiple qubits giving outcomes $\vec{a} = a_1a_2\cdots a_n$ where $a_i\{0,1\}$ for $i=1,\cdots, n$ is given by,
\bea
|\vec{a}\rangle \langle \vec{a}| = |a_1 \rangle\langle a_1| \otimes  |a_2 \rangle\langle a_2 |\otimes \cdots \otimes  | a_n \rangle \langle a_n |.  \nonumber
\eea
One can also rephrase a noiseless measurement as an estimation of $n$-qubit observables $Z_1\otimes Z_2 \otimes \cdots \otimes Z_n$ where $Z_j = |0\rangle \langle 0| -|1\rangle \langle 1|$ for all $j =1,\cdots, n$. 

In a realistic scenario, let $\Pi_{\vec{a}}$ denote a POVM giving outcomes $\vec{a}$. A measurement contains crosstalk noise if it cannot be decomposed into a product form, i.e.,
\bea
\mathrm{crosstalk~noise:~~}\Pi_{\vec{a}} \neq \Pi_{a_1} \otimes \cdots \otimes \Pi_{a_n}. \label{eq:cro}
\eea
The right-hand-side above describes a noisy POVM element that does not contain crosstalk noise.

\begin{figure}
    \centering
    \includegraphics[width=\linewidth]{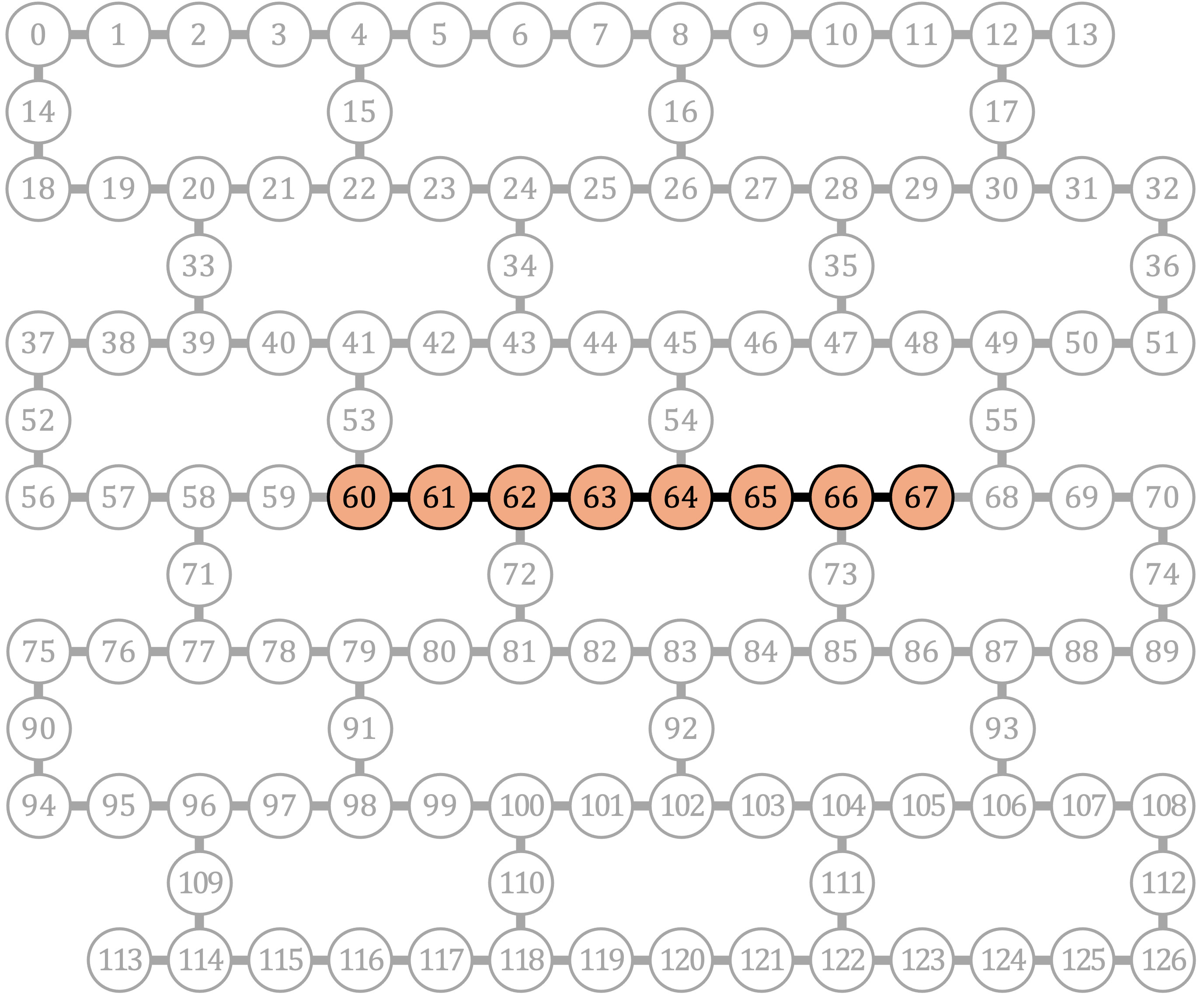}
    \caption{ A currently available quantum hardware contains noise. Throughout, we consider cloud-based quantum computing services and the device `ibm\_brisbane' with $127$ qubits. We apply the detection of measurement crosstalk and its mitigation on $8$ qubits, labeled $60$ to $67$. In Table \ref{tab:crosstalk1}, we show the presence of measurement crosstalk. In Fig. \ref{fig:diagram}, we demonstrate the mitigation protocols on noisy measurements over the qubits. } 
    \label{fig:ibm127b}
\end{figure}

\subsection{Quantification of measurement noise and crosstalk}

To quantify noisy POVM elements, it is generic to estimate their distinguishability with noiseless instances, 
\bea
\| \Pi_{\vec{a}} - |\vec{a}\rangle \langle \vec{a}| \|, 
\eea
for a relevant distance measure such as $1$-norm. We here restrict the consideration to the chosen basis, see noiseless measurement in Eq. (\ref{eq:nno1}). The quantification of a noisy POVM element from its ideal realization is computed by the fidelity, 
\bea
F( \Pi_{ \vec{a}} ) = \langle \vec{a} | \Pi_{\vec{a}}  |\vec{a} \rangle.  \label{eq:fidpovm}
\eea 
It is clear that if $\Pi_{ \vec{a} }$ forms a noiseless measurement, we have $F(\Pi_{ \vec{a} })=1$ for all $\vec{a}$. Otherwise, we have $F(\Pi_{ \vec{a} })<1$.

Although the above quantification is commonly used to characterize error, the resulting estimate does not capture only single-measurement error but also crosstalk error. Measurement crosstalk noise can be identified when a POVM element of a joint measurement cannot be characterized by separated measurement devices.

For simplicity, let us consider measurements on two qubits denoted by $S_1$ and $S_2$. QDT is performed on a two-qubit joint measurement and verifies a POVM element, denoted by  $\Pi_{a_1a_2}$. Let us also consider QDT performed on individual measurements with the corresponding POVM elements, denoted by $\Pi_{a_1}$ and $\Pi_{a_2}$. One can conclude that measurement crosstalk exists when a POVM element $\Pi_{a_1a_2}$ is not a product of them,
\bea
\Pi_{a_1a_2} \neq \Pi_{a_1 }  \otimes \Pi_{ a_2}.  \label{eq:crosdef}
\eea
When crosstalk noise is present, measurement outcomes of individual qubits cannot be independent. We introduce the gap as follows,  
\bea
\Delta_{a_1a_2 }^{(S_1S_2)} = \Pi_{a_1a_2} - \Pi_{a_1 }  \otimes \Pi_{ a_2}, \label{eq:del}
\eea
and note that 
\bea
 \tr_{S_1}\Delta_{a_1a_2}^{(S_1S_2)} = \tr_{S_2} \Delta_{a_1a_2}^{(S_1S_2)} = 0
.\nonumber 
\eea
Since measurement crosstalk noise concerns correlations between measurements of individual qubits, we introduce a measure for the crosstalk \cite{9645257},
\bea
\mathcal{C}(\Pi_{a_1a_2}) = \min_{M_{a_1}, M_{ a_2} \geq 0} \| \Pi_{a_1 a_2} - M_{a_1} \otimes M_{ a_2} \|_2 \label{eq:measure}
\eea
where the minimization runs over all POVM elements $M_{a_1}$ and $M_{a_2}$. We have used $\Vert X\Vert_2:=\sigma_{max}(X)$ where $\sigma_{max}(X)$ is the maximal singular value of $X$. Note that $\mathcal{C} (\Pi_{a_1a_2}) =0$ if and only if a POVM element on two qubits can be factorized. When computing the crosstalk measure, one can exploit a Kraus operator parameterized as follows,
$$K_{i} = \begin{pmatrix}
 k_1 & 0\\
 k_3+ik_4 & k_2
\end{pmatrix}$$
for the optimization of POVM elements $M_{i}=K^\dagger_{i} K_{i}$.

\begin{table}[h]
    \centering
    \caption{Computation of the crosstalk measure} 
    \label{tab:crosstalk1}
    \begin{tabular}{|c|c|c|c|}
        \hline
         &  &  &  \\
        ~Qubits~ & $F(\Pi_{00})$ & $~  \mathcal{C}(\Pi_{00})~ $  &  $\mathcal{C}(U^\dagger\Pi_{00} U)$\\
        \hline 
        67,66 & 0.890 & 0.093  & 0.014  \\
        67,63 & 0.895 & 0.122  & 0.030  \\
        67,60 & 0.886 & 0.303  & 0.307  \\
        \hline
    \end{tabular}
\end{table}

In Table \ref{tab:crosstalk1}, the measure for measurement crosstalk noise is computed in the fake backend `FakeBrisbane' that mimics real QPU `ibm\_brisbane', see Fig. \ref{fig:ibm127b}. QDT is performed on multiple qubits, three pairs of qubits $(67,66)$, $(67,63)$, and $(67,60)$. A two-qubit state $|00\rangle$ is prepared, and the probability of having outcomes $00$ is collected. The probabilities are $0.890$, $0.895$, and $0.886$, respectively, for the three pairs. The crosstalk noise $C(\Pi_{00})$ is computed, which can also decrease by quantum preprocessing that applies a local unitary $U$, i.e., $\mathcal{C}(U^\dagger\Pi_{00} U) < \mathcal{C}( \Pi_{00}  )$, where an optimal local unitary can be constructed, see Subsection \ref{subsec:protocol1}.   

\section{Measurement-error mitigation protocol}
\label{sec:p}

In this section, we present protocols for suppressing measurement errors on multiple qubits. The protocols contain two steps, one a quantum preprocessing applying local unitaries, and the other a classical postprocessing that adjusts the outcome statistics. To construct pre- and post-processings, it is required to perform QDT in advance to have noisy measurements characterized in terms of POVM elements.

\subsection{The framework }

As a prerequisite, let us suppose that QDT is performed on $n$-qubit measurements. A POVM element 
\bea
\Pi_{\vec{a}}~~\mathrm{for~giving~outcomes~} \vec{a} =a_1\cdots a_n \label{eq:povm}
\eea
is thus provided, from which we construct local unitaries for preprocessing and classical postprocessing. We refer to two decompositions of a POVM element, one we call preferred-basis decomposition (PBD), which is for Protocol 1, and the other eigendecomposition (ED) for Protocol 2.

The goal is clearly to approximate an outcome probability from noisy measurements into that of a noiseless one. A mitigation protocol, denoted by $\mathrm{MIT}$, signifies a method to minimize a distance in the following: for a $n$-qubit state $\rho$, 
\bea
\min_{ \mathrm{MIT}}~ \left| ~ [\mathrm{MIT} ( \rho, \Pi_{\vec{a}} ) - \tr[\rho |\vec{a}\rangle\langle \vec{a}|] ~ \right|\nonumber
\eea
where $\tr[\rho \proj{\vec{a}}]$ denotes a probability of $\vec{a}$ that can be obtained by a noiseless measurement.

\begin{figure}[t]
    \centering
    \includegraphics[width=0.9\linewidth]{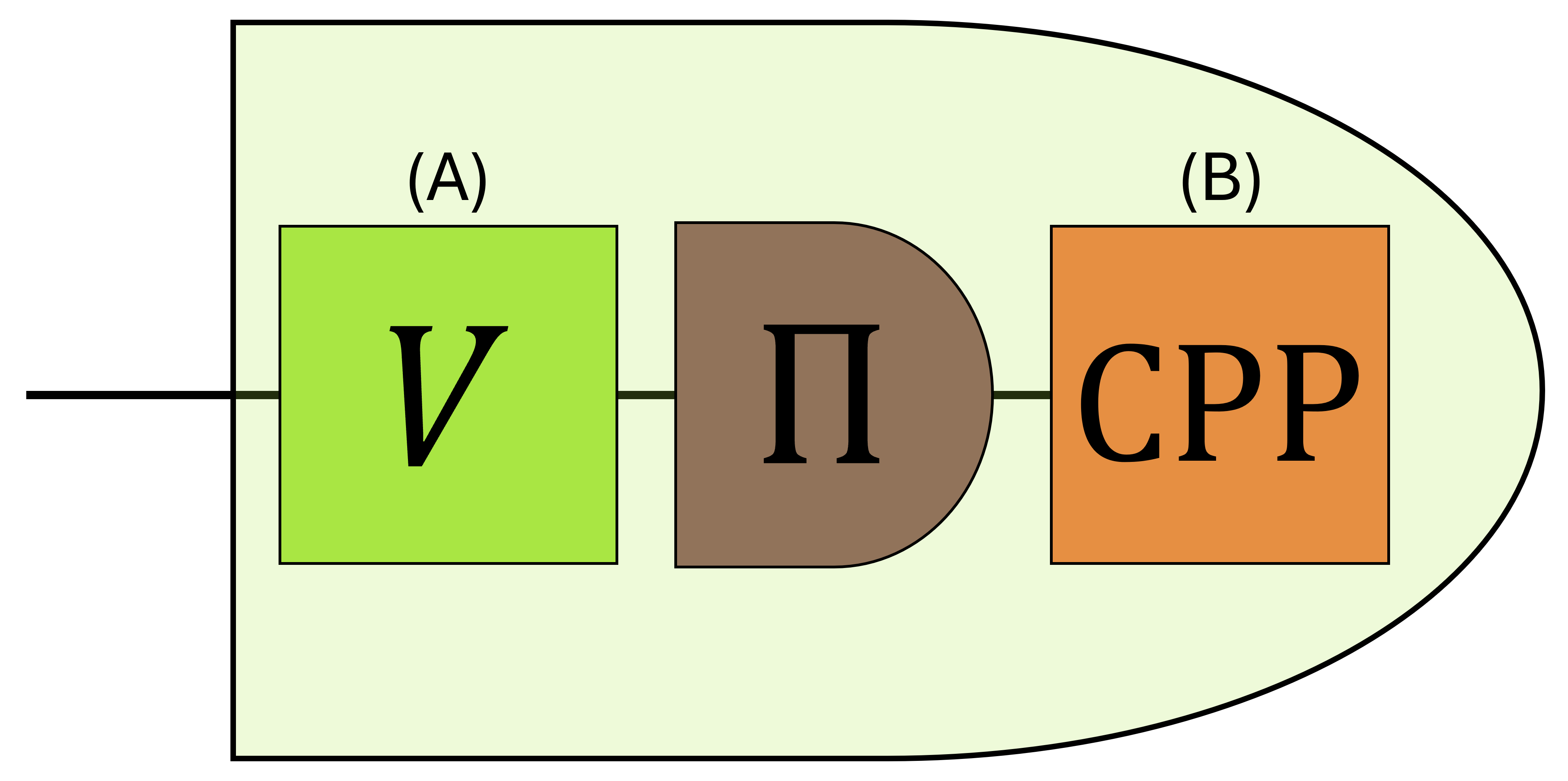}
    \caption{ Protocols for mitigating measurement errors throughout contain two stages, quantum preprocessing that applies local unitaries before a measurement, and classical postprocessing that manipulates probabilities of measurement outcomes. (A) The preprocessing realizes local unitaries constructed from a decomposition of a POVM element. Application of a unitary $V$ before a measurement can be equivalently formulated as a mapping for a POVM element, $\Pi \mapsto V^{\dagger} \Pi V$. (B) Classical postprocessing approximates a probability distribution from a noiseless measurement.}
    \label{fig:pi}
\end{figure}

\begin{figure*}[t]
		\centering
        \includegraphics[angle=0, width=\textwidth]{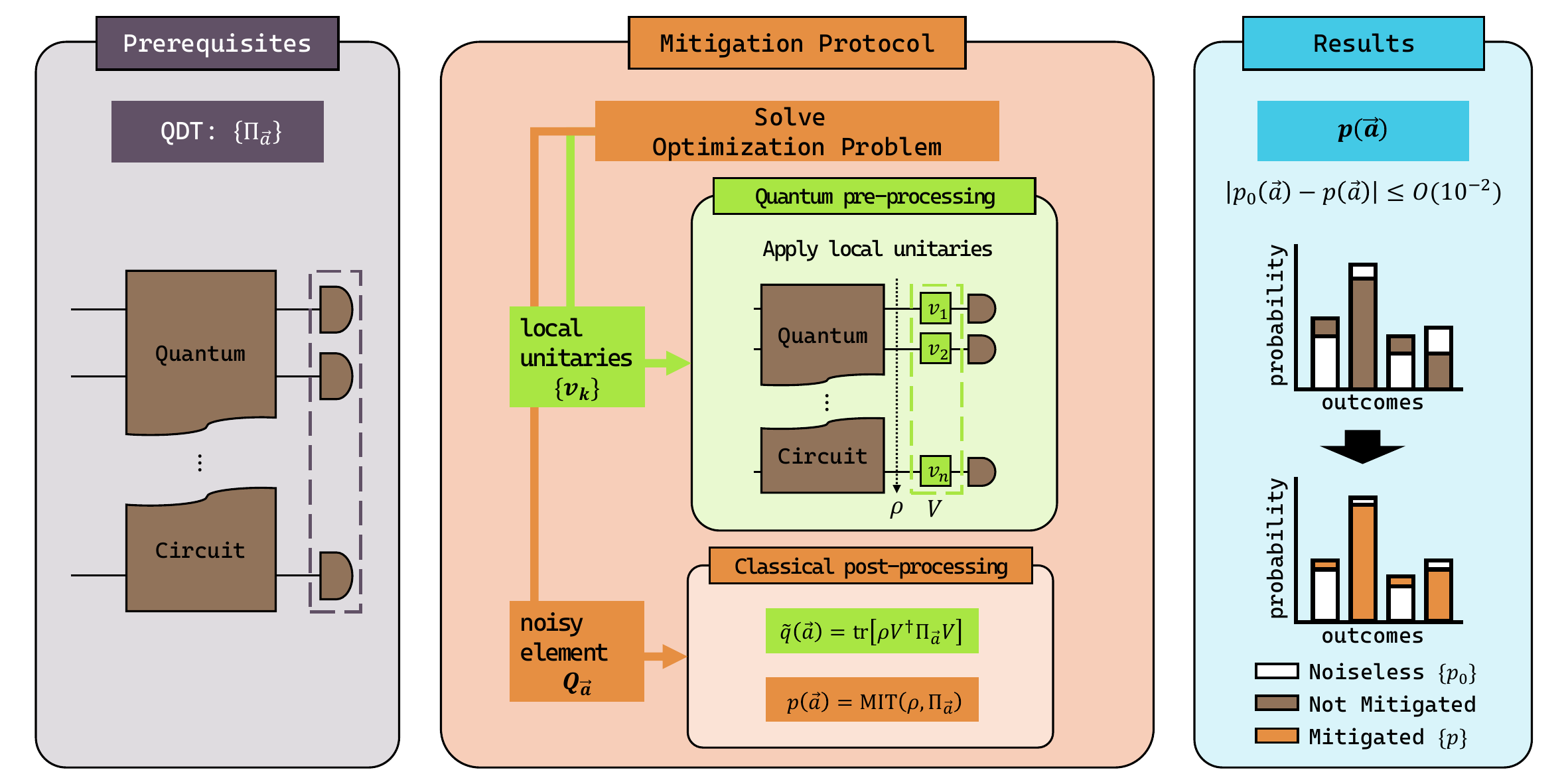}
		\caption{  The framework of the protocol of measurement error mitigation is summarized. (1) Prerequisites: QDT on a noisy measurement is performed, and POVM elements are provided. (2) The mitigation protocol has two steps, the QPP and the CPP, both of which are constructed from a POVM element verified by QDT. By analyzing decompositions of a POVM, a local unitary $V = v_1\otimes \cdots v_n$ is obtained, see Eq. (\ref{eq:v}), and Protocols 1 and 2 are constructed accordingly, see Eqs. (\ref{eq:dcomq}) and (\ref{eq:minv}). (3) The mitigation protocol approximates probabilities of noiseless measurements. An error in the approximation is up to $O(10^{-2})$.} 
        \label{fig:diagram}
\end{figure*}

We devise a protocol $\mathrm{MIT}$ such that it contains local unitaries that have lower error rates than measurements. For instance, superconducting qubits show a measurement error rate ten times higher than a single-qubit gate. Thus, the motivation of the mitigation protocol is to consume single-qubit gates, with much fewer errors and more feasible with present-day quantum technologies, to suppress measurement errors. The flow of mitigation protocols we propose is summarized in Fig. \ref{fig:diagram}.

\subsection{ Protocol 1 }
\label{subsec:protocol1}

We here exploit a PBD of a POVM element as follows, 
\bea
\Pi_{\vec{a}} = |\vec{a}\rangle \langle \vec{a}| +    P_{\vec{a}} \label{eq:dcom}
\eea 
where $| \vec{a}\rangle$ is a {\it preferred basis} and $P_{\vec{a}}$ is a Hermitian operator corresponding to the difference between noisy and noiseless POVM elements. 

{\it The quantum preprocessing (QPP)} is to exploit a local unitary transform $V$ before a noisy measurement
\bea
V : = v_1\otimes v_2\otimes \cdots \otimes v_n \label{eq:v} 
\eea
where $v_i$ denotes a local unitary for the $i$-th qubit. An application of a unitary $V$ before a noisy measurement takes place, see Fig. \ref{fig:pi}, can be seen equivalently as a transformation of a POVM element as follows, 
\bea
V^{\dagger} \Pi_{\vec{a}} V 
= |\vec{a}\rangle \langle \vec{a}| +     Q_{\vec{a}} \label{eq:dcomq}
\eea
with the preferred basis $|\vec{a}\rangle$ and $Q_{\vec{a}}$ is a Hermitian operator corresponding to a difference between two operators $V^{\dagger} \Pi_{\vec{a}} V $ and $|\vec{a}\rangle \langle \vec{a}|$, see also Eq. (\ref{eq:dcom}).

\begin{figure}[t]
		\centering
        \includegraphics[angle=0, width=0.45\textwidth]{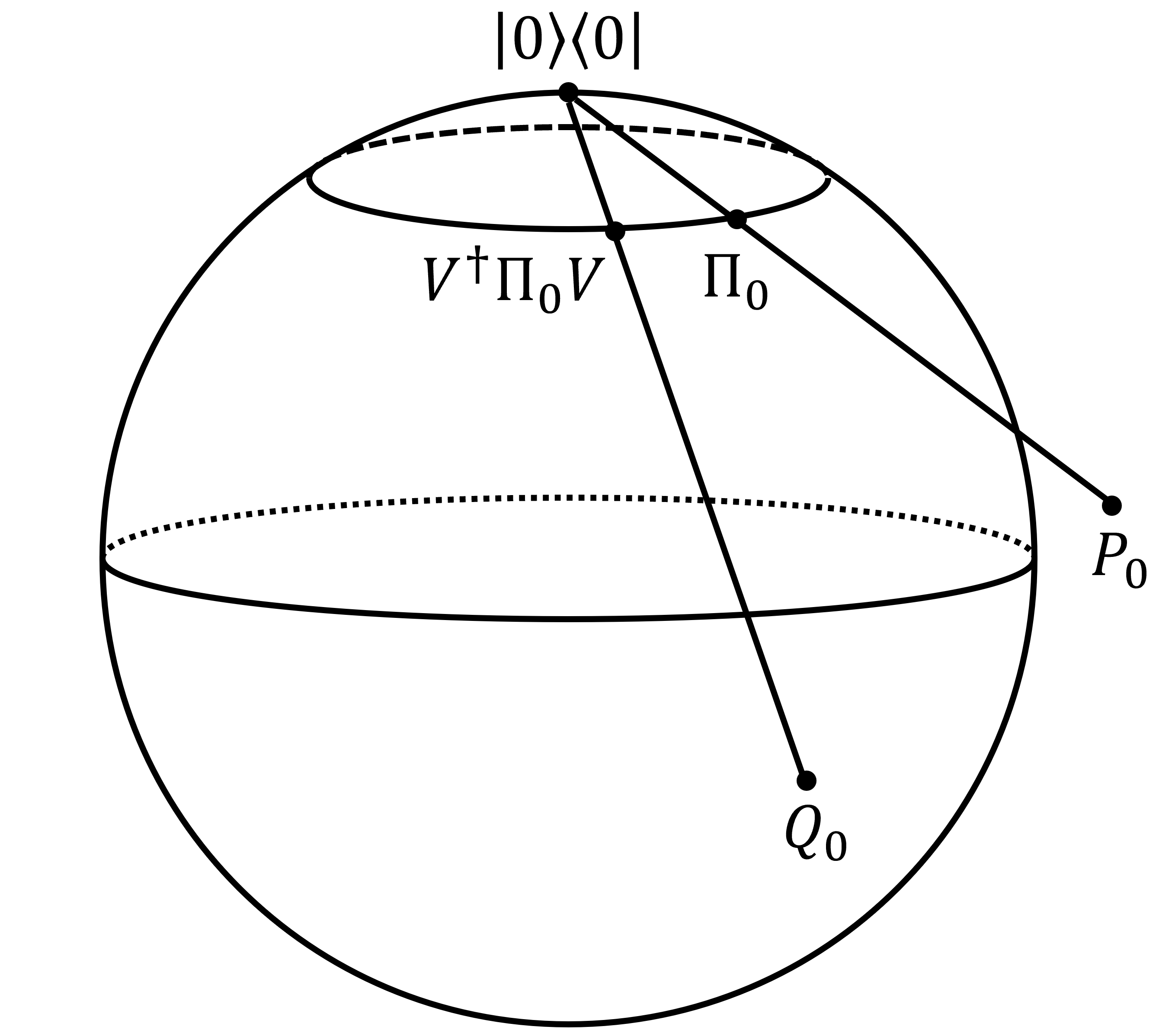}
		\caption{ Consider a noisy POVM element $\Pi_0$, deviated from an ideal one $|0\rangle \langle 0|$. It can be a convex combination of the ideal one $|0\rangle \langle 0|$ with a Hermitian operator $P_0$, see Eq. \eqref{eq:dcom}. It may be unitarily transformed to $V^{\dagger} \Pi_0 V$; this transformation can be realized by placing a unitary $V$ before a measurement, see Fig. \ref{fig:pi}. Once it is transformed, the resulting one $V^{\dagger} \Pi_0 V$ can be a convex combination of the preferred basis $|0\rangle \langle 0|$ with some other Hermitian operator $Q_0$, see  Eq. (\ref{eq:dcomq}).} 
        \label{fig:pi2}
\end{figure}

The reason to exploit a local unitary $V$ is to increase the fidelity of a noisy POVM element as follows, $$F(\Pi_{\vec{a}})=\bra{\vec{a}}\Pi_{\vec{a}}\ket{\vec{a}}.$$
Then, a unitary transform $V$ that increases the fidelity is thought, 
\bea
\max_V~ \langle \vec{a} | V^{\dagger} \Pi_{\vec{a}} V  |\vec{a} \rangle   \geq  \langle \vec{a} | \Pi_{\vec{a}}  |\vec{a} \rangle.  \label{eq:rotation}
\eea
This also means that $\langle \vec{a} |    Q_{\vec{a}} | \vec{a}\rangle \leq  \langle \vec{a} |   P_{\vec{a}} | \vec{a}\rangle$, see Fig. \ref{fig:pi2}.

\textit{The classical postprocessing (CPP)} can be devised once a measurement is performed with a POVM element in Eq. (\ref{eq:dcomq}). Let us write by
\bea
\widetilde{q}(\vec{a}) = \tr[\rho V^{\dagger} \Pi_{\vec{a}} V]~~\mathrm{and}~~ p_0 (\vec{a}) = \tr [\rho | \vec{a} \rangle \langle \vec{a} | ], \eea
and it holds that 
\bea
\widetilde{q}(\vec{a}) = p_0 (\vec{a}) +  \tr[Q_{\vec{a}} \rho] \label{eq:qq}
\eea
from which a probability from a noiseless measurement is given by
\bea
p_0 (\vec{a}) = \widetilde{q}(\vec{a}) -   \tr [Q_{\vec{a}} \rho]. \label{eq:p0}
\eea
Note that $\widetilde{q}(\vec{a})$ is given from measurements but $\tr [Q_{\vec{a}} \rho]$ is not known; while $Q_{\vec{a}}$ has been verified by QDT of a noisy measurement, $\tr [Q_{\vec{a}} \rho]$ cannot be computed immediately for an unknown state $\rho$. As it is bounded by $(Q_{\vec{a}})_{\min}$ and $(Q_{\vec{a}})_{\max}$, we make an approximation as its average:
\bea
p (\vec{a}) &=& \widetilde{q}(\vec{a}) - \bar{Q}_{\vec{a}} \label{eq:rec} \\
\mathrm{where} && \bar{Q}_{\vec{a}} = \frac{1}{2} \left(  (Q_{\vec{a}})_{\max} + (Q_{\vec{a}})_{\min}   \right).  \nonumber 
\eea
Here, the maximum and minimum eigenvalues of $Q_{\vec{a}}$ are written by:
\bea
(Q_{\vec{a}})_{\max}&:=&\max_{\rho} \tr[Q_{\vec{a}}\rho]\nonumber\\
\text{and}~~(Q_{\vec{a}})_{\min}&:=&\min_{\rho} \tr[Q_{\vec{a}}\rho].\nonumber
\eea

Finally, as $(Q_{\vec{a}})_{\min}$ and $(Q_{\vec{a}})_{\max}$ are closer, the resulting probability $p(\vec{a})$ makes a better approximation to a noiseless one $p_0(\vec{a})$. For this reason, we find a local unitary in QPP such that the difference is minimized, 
\bea
&& \min_V ~  \left| (Q_{\vec{a}})_{\max} - (Q_{\vec{a}})_{\min}\right|\label{eq:minp}
\eea 
where $Q_{\vec{a}} = V^{\dagger}\Pi_{\vec{a}} V - |\vec{a}\rangle \langle \vec{a}| $. Protocol 1 is summarized in $\mathrm{Algorithm}$ \ref{alg:pbd}.

Let us now show that Protocol 1 derives a resulting probability distribution which is $\epsilon$-close to a noiseless probability, i.e., $\left| p(\vec{a}) - p_{0}(\vec{a}) \right| = O(10^{-2})$. First, we note that $|\bar{Q}_{\vec{a}}| = O(10^{-2})$ since it is a product of decimals 
\bea
|\bar{Q}_{\vec{a}}| = | \tr Q_{\vec{a}} | \times \left| \frac{ \bar{Q}_{\vec{a}} }{\tr Q_{\vec{a}}} \right| = O(10^{-2}) \label{eq:barq} 
\eea
where $|\tr Q_{\vec{a}} | = |1-\tr \Pi_{\vec{a}}|<1$ from Eq. (\ref{eq:dcomq}) and, clearly, $| { \bar{Q}_{\vec{a}} }/ {\tr Q_{\vec{a}}}| <1$.  The difference between a noiseless probability and a mitigated one can be found from Eqs. (\ref{eq:p0}) and (\ref{eq:rec}), and we compute,
\bea
\left| p(\vec{a}) - p_{0}(\vec{a}) \right| &=&  \left| \tr Q_{\vec{a}}\rho -  \bar{Q}_{\vec{a}}  \right|. \label{eq:right}
\eea
The right-hand side can be rewritten as, 
\bea
&& |\tr Q_{\vec{a}} | \times  \left| \tr \left( \frac{Q_{\vec{a}}}{ \tr Q_{\vec{a}} }  \rho\right) -   \frac{ \bar{Q}_{\vec{a}} }{\tr Q_{\vec{a}}} \right| \nonumber 
\eea
where $|\tr Q_{\vec{a}} | = |1-\tr \Pi_{\vec{a}}|<1$ from Eq. (\ref{eq:dcomq}), and also clearly the next term is less than $1$. Hence, we have an estimate $\left| p(\vec{a}) - p_{0}(\vec{a}) \right|  = O(10^{-2})$. We remark that the rate of suppression is independent of the number of qubits; hence, for any $n$-qubit  measurement, Procotol 1 can suppress a measurement error rate up to a percentage level. 




\begin{algorithm}[t]
\caption{Protocol 1}\label{alg:pbd}

    \KwData{POVM element $\Pi_{\vec{a}}$ from QDT,\\
            ~~~~~~~~~ noiseless projective measurement $\proj{\vec{a}}$, \\
            ~~~~~~~~~ target outcome $\vec{a}$.}
            
    \KwResult{Mitigated probability $p(\vec{a})$} 

    ~\\
    {\bf Quantum Preprocessing:}\\
    \begin{enumerate}
        \item[1.] Find the decomposition: $\Pi_{\vec{a}}=\proj{\vec{a}} + P_{\vec{a}}$\\ 
        \item[2.] Find a local unitary $V$ such that 
        $$\min_{V} \left| (Q_{\vec{a}})_{\max}-(Q_{\vec{a}})_{\min} \right|\quad\mathrm{s.t.}\quad Q_{\vec{a}}=V^\dagger \Pi_{\vec{a}} V -\proj{\vec{a}} $$
        \item[3.] Return QPP $V^\dagger\Pi_{\vec{a}} V=\proj{\vec{a}} + Q_{\vec{a}}$
    \end{enumerate}

    {\bf Classical Postprocessing:}\\
    For a given state $\rho$, $\widetilde{q}(\vec{a})=p_0(\vec{a})+ \tr\left[ \rho Q_{\vec{a}} \right]$
    \begin{enumerate}
        \item[1.] Find the maximum and minimum eigenvalues of $Q_{\vec{a}}$, $(Q_{\vec{a}})_{\max}$ and $(Q_{\vec{a}})_{\min}$\\
        \item[2.] Replace $\tr\left[ \rho Q_{\vec{a}} \right]$ with \\$\bar{Q}_{\vec{a}}=\left((Q_{\vec{a}})_{\max}+(Q_{\vec{a}})_{\min}\right)/2$\\
        \item[3.] Return CPP applied probability of an outcome $\vec{a}$
        $$p(\vec{a})=\widetilde{q}(\vec{a})-\bar{Q}_{\vec{a}}$$
    \end{enumerate}
\end{algorithm}

\subsection{ Protocol 2  }

Protocol 2 exploits an eigendecomposition (ED) of a POVM element as follows,
\bea
    \Pi_{\vec{a}} &=&\sum_{i=1}^d \alpha_{i}^{[\vec{a}]} \proj{\alpha_{i}^{[\vec{a}]} }, \nonumber\\
    &=& \alpha_{i}^{[\vec{a}]}  \proj{\alpha_{i}^{[\vec{a}]} } + P_{\vec{a}} \label{eq:edcom}
\eea
where $\{\alpha_{i}^{[\vec{a}]}  \}$ are eigenvalues which are sorted in descending order; $\alpha_{1}^{[\vec{a}]} \ge \alpha_{i}^{[\vec{a}]}  \ge \cdots \ge \alpha_{d}^{[\vec{a}]}  $ and thus $P_{\vec{a}}=\sum_{i=2}^d \alpha_{i}^{[\vec{a}]}   \proj{ \alpha_{i}^{[\vec{a}]}  }$. Protocol 2 also contains the QPP and the CPP, where the QPP finds a local unitary $V$ such that 
\bea
\max_{V}~\langle \vec{a} |V | \alpha_{1}^{[\vec{a}]}  \rangle. \label{eq:minv}
\eea
Thus, a local unitary $V$ realizes a best approximation to a rotation of a basis $| \alpha_{1}^{[\vec{a}]}  \rangle$ to a desired one $|a\rangle$. 

Once an optimal local unitary $V$ is verified, the next steps of the QPP and the CPP are straightforward. The QPP applies a local unitary $V$ to realize a transformation of a POVM element,
\bea
\Pi_{\vec{a}} \mapsto V^{\dagger} \Pi_{\vec{a}} V = \alpha_{1}^{[\vec{a}]}  V^{\dagger} \proj{ \alpha_{1}^{[\vec{a}]}  }V + V^{\dagger} P_{\vec{a}} V  \nonumber
\eea
Note that the eigenvalues of $P_{\vec{a}}$ remain the same under a local unitary $V$. A probability of measurement outcomes is bounded above and below as follows, 
\bea
\frac{1}{\alpha_{1}^{[\vec{a}]} } (\widetilde{q}(\vec{a}) -   ( P_{\vec{a}})_{\max} ) \leq p_0(\vec{a}) \leq \frac{1}{\alpha_{1}^{[\vec{a}]} } (\widetilde{q}(\vec{a}) -  (P_{\vec{a}})_{\min}) \label{eq:edbnd}
\eea
where we have written
\bea
(P_{\vec{a}})_{\min} & :=& \min_\rho \tr[P_{\vec{a}} \rho]  \nonumber\\
\mathrm{and}~~(P_{\vec{a}})_{\max} & :=& \max_\rho \tr[P_{\vec{a}} \rho]. \nonumber
\eea
It is straightforward to derive the upper bound from Eq. (\ref{eq:edbnd}),
\bea
&& | p_0(\vec{a}) - p(\vec{a}) |\leq \frac{1}{ \alpha_{1}^{[\vec{a}]}  }\Delta \bar{P}_{\vec{a}},  \nonumber \\
\mathrm{where}&& \Delta\bar{P}_{\vec{a}} = \frac{1}{2} ((P_{\vec{a}})_{\max} - (P_{\vec{a}})_{\min}).
\eea
Protocol 2 is summarized in $\mathrm{Algorithm}$ \ref{alg:evd}. Following the lines in Eq. (\ref{eq:right}), we have $\Delta\bar{P}_{\vec{a}} = O(10^{-2})$ and thus the upper bound of the error, $\Delta\bar{P}_{\vec{a}} / \alpha_{1}^{[\vec{a}]}   $, is also in the order of $10^{-2}$.

\begin{algorithm}[t]
\caption{Protocol 2}\label{alg:evd}

    \KwData{Noisy POVM element $\Pi_{\vec{a}}$,\\
            ~~~~~~~~~ noiseless projective measurement $\proj{\vec{a}}$, \\
            ~~~~~~~~~ target outcome $\vec{a}$.}
            
    \KwResult{Mitigated probability $p(\vec{a})$} 

    ~\\
    {\bf Quantum Preprocessing:}\\
    \begin{enumerate}
        \item[1.] Find the eigendecomposition of $\Pi_{\vec{a}}$, $\Pi_{\vec{a}}= \alpha_{1}^{[\vec{a}]}   \proj{ \alpha_{1}^{[\vec{a}]}  } + P_{\vec{a}}$ where $\alpha_{1}^{[\vec{a}]}   = \max \mathrm{eig(\Pi_{\vec{a}})}$. 
        \item[2.] Find a local unitary $V$ such that  $\ket{\vec{a}}=V^\dagger\ket{\alpha_{1}^{[\vec{a}]} }$
        $$\max_{V} \left| \bra{  \alpha_{1}^{[\vec{a}]}  }V\ket{\vec{a}} \right| ^2$$
        \item[3.] Return QPP
        $V^\dagger\Pi_{\vec{a}} V = \alpha_{1}^{[\vec{a}]}  \proj{\vec{a}} + V^\dagger P_{\vec{a}} V$
    \end{enumerate}

    {\bf Classical Postprocessing:}\\
    For a given state $\rho$, $\widetilde{q}(\vec{a}) = \alpha_{1}^{[\vec{a}]}  p_0(\vec{a})+ \tr\left[ \rho V^\dagger P_{\vec{a}} V \right]$
    \begin{enumerate}
        \item[1.] Find the maximum and minimum eigenvalues of $P_{\vec{a}}$, $(P_{\vec{a}})_{\max}$ and $(P_{\vec{a}})_{\min}$\\
        \item[2.] Replace $\tr\left[ \rho V^\dagger P_{\vec{a}} V \right]$ with \\
        $\bar{P}_{\vec{a}}= (( P_{\vec{a}})_{\max}+(P_{\vec{a}})_{\min} ) /2$\\
        \item[3.] Return CPP applied probability of an outcome $\vec{a}$
        $$p(\vec{a})=(\alpha_{1}^{[\vec{a}]} )^{-1}[\widetilde{q}(\vec{a})-\bar{P}_{\vec{a}}]$$
    \end{enumerate}
\end{algorithm}


\subsection{Application in a realistic scenario}

As a realistic scenario, the QPP and the CPP are implemented in the fake backend `FakeBrisbane' that mimics the real QPU `ibm\_brisbane', see Fig. \ref{fig:ibm127b}. In Table \ref{tab:example1}, qubits labeled $66$ and $67$ are considered. From the results of QDT, the fidelities of POVM elements are obtained $F(\Pi_0)$, which increase by a local unitary $U$. From decompositions in Eqs. (\ref{eq:dcom}) and (\ref{eq:dcomq}), we also compute $\Delta \bar{P}_0 =  ( (P_0)_{\max}-(P_0)_{\min} )/2 $ and $\Delta \bar{Q}_0 =   ((Q_0)_{\max}-(Q_0)_{\min} ) /2$. 


\begin{table}[h!]
    \centering
    \caption{ The effect of local unitaries on a POVM element}
    \label{tab:example1}
    \begin{tabular}{|c|c|c|c|c|}
        \hline
         &  &  &   &  \\
        ~~Qubit~~ & $F(\Pi_0)$ & ~$F(U^\dagger \Pi_0 U)$~ & $\Delta \bar{P}_0$ & $\Delta \bar{Q}_0 $\\
        \hline  
        66 & ~0.98169~  & 0.98170 & ~0.01966~ & ~0.01933~ \\ 
        67 & ~0.93220~  & 0.93221  & ~0.07015~ & ~0.07003~ \\ 
        \hline
    \end{tabular}
\end{table}


\section{Limitations on single-qubit readout mitigation} 
\label{sec:l}

We show here that measurement crosstalk errors cannot be mitigated by repeating a protocol designed to mitigate single-qubit measurement errors. Note that Protocols 1 and 2 apply to mitigating measurement errors of $n$ qubits in general. In the following, we apply Protocol 1 to mitigate single-qubit measurement errors. 

For convenience, let us consider measurements on two qubits and their outcomes $a_1 a_2$. By a mitigation protocol on each qubit, we have a resulting probability in the following form, see also Subsection \ref{subsec:protocol1},
\bea
p(a_1)p(a_2) = \left(\widetilde{q}(a_1) - \bar{Q}_{a_1}\right)  \left(\widetilde{q}(a_2) - \bar{Q}_{a_2} \right). ~~~\label{eq:ind}
\eea
When the mitigation protocol applies collectively, we exploit a local unitary $V$ leading to a decomposition as follows, 
\bea
V^{\dagger} \Pi_{a_1a_2} V = |a_1a_2\rangle \langle a_1a_2 | + Q_{a_1a_2} \label{eq:two}
\eea
with a local unitary $V = v_1 \otimes v_2$. From above, a probability distribution in a noiseless measurement can be seen as
\bea
p_0 (a_1a_2) = \widetilde{q} (a_1 a_2)  - \tr[ Q_{a_1a_2} \rho] \label{eq:noitwo} 
\eea
where $ \widetilde{q} (a_1 a_2) = \tr[ V^{\dagger} \Pi_{a_1a_2} V \rho]$. In the following, we show that probabilities obtained by mitigating measurement errors on individual qubits, see Eq. (\ref{eq:ind}) contain the effect of crosstalk noise.

\begin{figure}[t]
    \centering
    \includegraphics[width=\linewidth]{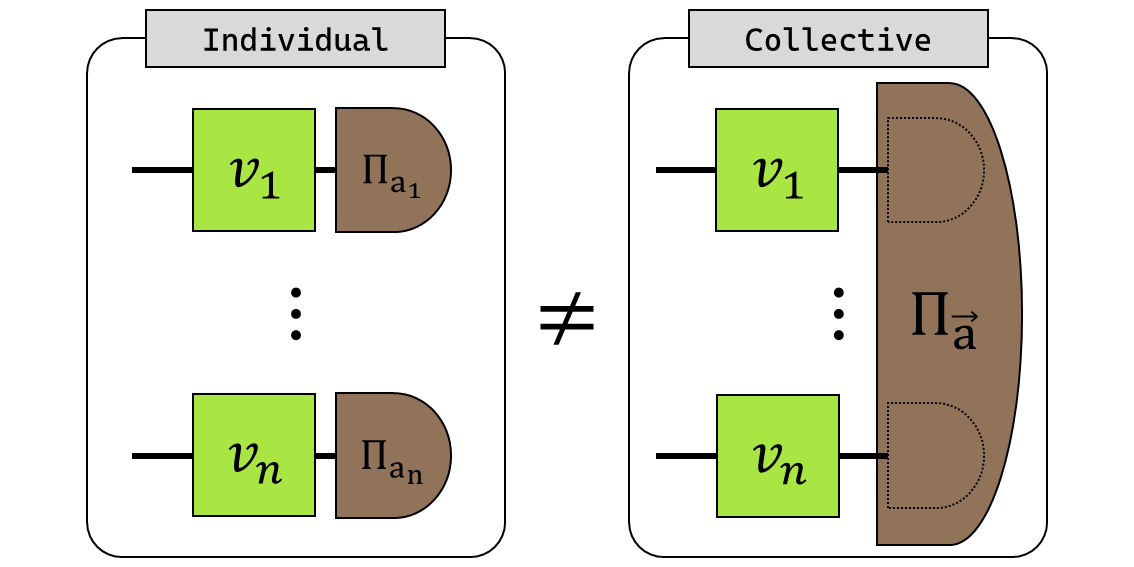}
    \caption{ Measurement error mitigation can be realized on individual qubits or on multiple qubits. The former does not deal with crosstalk noise, while the latter can suppress crosstalk errors. The effect of crosstalk noise remains if measurement error mitigation applies to individual qubits, see Eq. (\ref{eq:wash}).
    }    
    \label{fig:2}
\end{figure}

We compute how far the probabilities in Eq. (\ref{eq:ind}) from single-qubit error mitigation are from noiseless ones in Eq. (\ref{eq:noitwo}). Without loss of generality, we can consider a bipartite state $\rho=\rho_1 \otimes \rho_2 $ and show that
\bea
   p_0 (a_1a_2) - p (a_1) p(a_2)  = \tr[\Delta_{a_1a_2} V \rho V^{\dagger} ] + O(10^{-2}),~~ \label{eq:wash}
\eea
where $\Delta_{a_1a_2} = \Pi_{a_1a_2} - \Pi_{a_1}\otimes  \Pi_{a_2}$, see in Eq. (\ref{eq:cro}) defined for measurement crosstalk.

Recall that $|\bar{Q}_{a_1}|$, $|\bar{Q}_{a_2}|$, and $|\bar{Q}_{a_1 a_2}|$ are in a level of $10^{-2}$, see Eq. (\ref{eq:barq}). From Eq. (\ref{eq:ind}), we have for local unitaries $V=v_1\otimes v_2$ and bipartite state $\rho$
\bea
\widetilde{q}(a_1) \widetilde{q}(a_2) & = & \tr[v_{1}^{\dagger} \Pi_{a_1} v \rho_1] \tr[v_{2}^{\dagger} \Pi_{a_2} v \rho_2] \nonumber \\
& = & \tr  [V^{\dagger} \Pi_{a_1} \otimes \Pi_{a_2} V \rho]\nonumber 
\eea
where $\rho_i$ is a reduced state for system $i=1,2$ and $\Pi_{a_1} = \tr_{2} \Pi_{a_1a_2}$ and $\Pi_{a_2} = \tr_{1} \Pi_{a_1a_2}$. It follows that,
\bea
p_{0} (a_1 a_2) - \widetilde{q}(a_1) \widetilde{q}(a_2) = \tr[V^{\dagger} \Delta_{a_1a_2} V \rho ] + O(10^{-2})\nonumber
\eea
which proves Eq. (\ref{eq:wash}). Hence, errors resulting from crosstalk noise may be mitigated by a two-qubit {\it collective} mitigation protocol, not by {\it individual} mitigation protocol, see Fig. \ref{fig:2}. 

\subsubsection*{Analysis in a realistic scenario}

\begin{table}[h]
    \centering
    \caption{ Comparison: collective versus individual error mitigation } 
    \label{tab:example2}
    \begin{tabular}{|c|c|c|c|c|}
        \hline
         &  &   &  \\
        ~Qubits~ & $F(\Pi_{00})$ & $~ p(00)~$ & $~ p(0)p(0)~ $\\
        \hline 
        67,66 & 0.890  & 0.944 &  0.913  \\
        67,63 & 0.895  & 0.946  & 0.919  \\
        67,60 & 0.886  & 0.942 & 0.900  \\
        \hline
    \end{tabular}
\end{table}

The mitigation protocol applies to three pairs of qubits $(67,66)$, $(67,63)$, and $(67,60)$ in the fake backend `FakeBrisbane' that mimics real QPU `ibm\_brisbane'. We consider a state prepared in $|00\rangle$ and measurement outcomes $00$; the measurement fidelities $F(\Pi_{00})$ are given by $0.890$, $0.895$, and $0.886$. The mitigation protocol for single-qubit measurements corrects the noisy outcomes and improves up to $0.913$, $0.919$, and $0.900$, while leaving crosstalk errors uncorrected, see Eq. (\ref{eq:wash}). The mitigation protocol on two-qubit measurements, thereby correcting crosstalk errors, improves noisy probabilities up to $0.944$, $0.946$, and $0.942$, respectively. Hence, it is demonstrated that the mitigation protocol for two-qubit measurements collectively strictly outperforms the case of measurement error mitigation on single qubits. This also implies the significance of crosstalk noise on multiple qubits in realistic scenarios. 


\section{Applications to noisy quantum hardware}
\label{sec:a}

In this section, we apply the mitigation protocols to realistic devices, with `FakeBrisbane' that simulates the noise model of a realistic cloud quantum computing device of the \texttt{IBMQ}, `ibm\_brisbane'. In Fig. \ref{fig:ibm127b}, the configuration of the qubit array is presented in which $8$ qubits labeled from $60$ to $67$ are considered to conduct the experiment. The proof-of-principle demonstration is implemented on the single qubits, qubit pairs, and qubit triples. 

\subsection{Details of \texttt{IBMQ} experiment} 

Local unitaries on $n$ qubits are denoted by $ V=\bigotimes_{i=1}^n v_i$ where $v_i$ for $i=1,\cdots, n$ is a single-qubit unitary. A single-qubit unitary can be parametrized by $(\theta,\phi, \lambda)$,
\bea\label{eq:uni}
v(\theta,\phi,\lambda)=\begin{pmatrix}
    \cos{\frac{\theta}{2}} & -e^{i\lambda}\sin{\frac{\theta}{2}} \\
    e^{i\phi}\sin{\frac{\theta}{2}} & e^{i(\phi+\lambda)}\cos{\frac{\theta}{2}}
\end{pmatrix}.
\eea
QDT is performed on single-, two-, and three-qubit measurements. For instance, POVM elements $\Pi_0$ of single-qubit measurements, giving outcome $0$, for qubits labeled $66$ and $67$ are obtained:
$$\Pi^{(66)}_0= 10^{-3} \times \begin{pmatrix}
    982 &  2- 6 i \\
   2  + 6i & 19
\end{pmatrix}$$
and
$$\Pi^{(67)}_0= 10^{-3} \times \begin{pmatrix}
    932 & -2 + 3 i \\
   - 2 - 3 i & 72 
\end{pmatrix}.$$
In Table \ref{tab:example1}, their fidelities are listed as well as the improvement by local unitaries and properties of $P$ and $Q$ operators of Protocol 1.


QDT is also performed on two qubits labeled $67$ and $66$ and reconstructed a POVM element
\bea
\Pi^{(67,66)}_{00}= 10^{-3}\times \begin{pmatrix}
   890  & -1  +   5 i & 2 -   4 i & 1  -  2 i\\
   * & 12 & -1  & 1   \\
   * & * & 93  & 0 \\
   * & * & * & 3
\end{pmatrix} \label{eq:6766}
\eea
from which we have the probability of having outcomes $00$ is $0.890$. We also compute the crosstalk measure in Eq. (\ref{eq:measure}) and obtain $0.093$. Note that a noiseless POVM element is $\mathrm{diag}[1,0,0,0]$.

\subsection{Experimental results from IBM demonstration}

\begin{figure}[t]
    \centering
    \includegraphics[width=\linewidth]{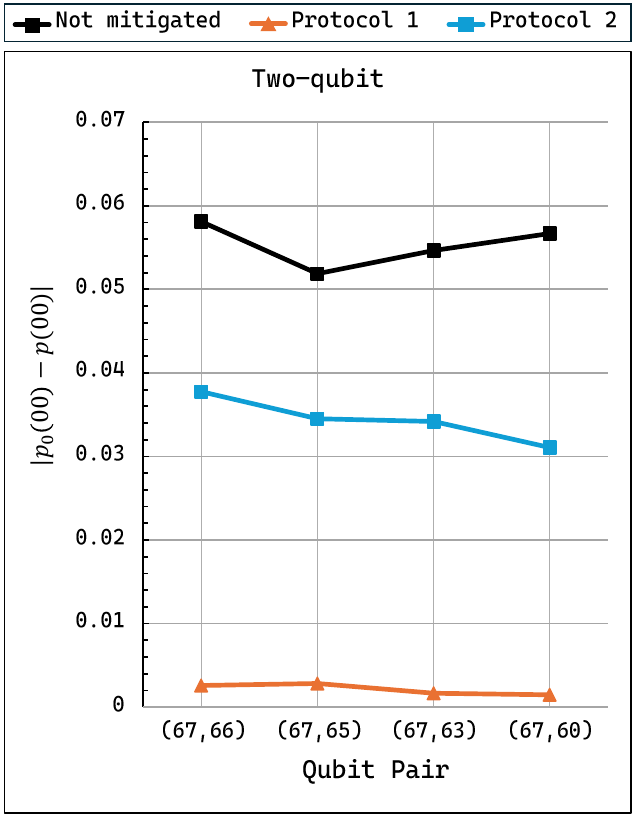}
    \caption{ The gap between probabilities of noiseless and noisy measurements for outcome $00$ is shown, which is between $0.05$ and $0.06$ (black). Protocol 2 suppresses the error rates so that the gap is bounded by $0.04$ (blue). Protocol 1 reduces the error rate with the gap up to $0.002$ (orange).  }
    \label{fig:comps0}
\end{figure}

We here apply Protocols 1 and 2 to mitigate measurement errors on multiple qubits. We also consider two protocols in the following scenarios: (i) mitigation of noisy measurements for a specific outcome and (ii) mitigation of noisy measurements for all outcomes on average. Both scenarios have distinct applications in quantum information processing. For instance, protocols such as quantum entanglement distillation or static error purification~\cite{kim2025protocolpurifyingnoisypreparation} rely only on specific outcomes. 


\subsubsection*{Mitigation of noisy measurement outcomes $00$} 

We consider noisy measurements on qubit pairs $(67,66)$, $(67,65)$, $(67,63)$, and $(67,60)$. QDT is performed on POVM elements giving outcome $00$. We apply Protocol 1 for which local unitaries $v_1$ and $v_2$ for the QPP are obtained, where parameters $(\theta, \phi, \lambda)$, see Eq. (\ref{eq:uni}), are given by
\bea
v_1:~(\theta_1,\phi_1,\lambda_1) &= & (0.0035\pi,0.3199\pi,-0.8831\pi) ~\mathrm{and}\nonumber \\
v_2:~(\theta_2,\phi_2,\lambda_2) &=& (-0.0039\pi,0.4221\pi,0.7015\pi). \label{eq:vv} 
\eea 
The results of measurement error mitigation by Protocols 1 and 2 are shown in Fig. \ref{fig:comps0}. In all cases, two protocols successfully suppress measurement error rates. In particular, some of the results are summarized in Table \ref{tab:example2} for qubits $66$ and $67$, the fidelity is improved from $0.890$ to $0.944$. Moreover, the crosstalk measure was reduced to $0.014$, see Table \ref{tab:crosstalk1}. The results for the other pairs of qubits, $(67,63)$ and $(67,60)$, are presented in Table \ref{tab:example2}.

\begin{table} [t]
    \centering 
    \caption{Error mitigation on three-qubit measurements}
    \label{tab:example3}
    \begin{tabular}{|c|c|c|c|}
        \hline
        &   &   &  \\
        Qubits & $F(\Pi_{000})$  & $~p(000)~$  &  ~ $\mathcal{C}(\Pi_{000})$ ~ \\
        \hline 
        ~67,66,65~ & ~0.875~ & ~0.938~ & 0.136 \\
        ~67,65,63~ & ~0.884~ & ~0.942~ & 0.092 \\
        ~67,63,61~ & ~0.879~ & ~0.939~ & 0.090 \\
        \hline
    \end{tabular} 
\end{table}
QDT of measurements on triples of qubits $(67,66,65)$, $(67,65,63)$, and $(67,63,61)$ are also performed. In Table \ref{tab:example3}, the measurement fidelities are summarized for a state $|000\rangle$ and a measurement outcome $000$: $0.875$, $0.884$, and $0.879$ for the triples, respectively. The mitigation protocol improves the probability up to $0.938$, $0.942$, and $0.939$.

\subsubsection*{Mitigation of noisy measurement outcomes  on average} Let us consider the scenario (ii) in which measurement error mitigation on all outcomes are attempted. For protocols, we construct QPPs by choosing optimal local unitaries $V$ obtained in the following, 
\bea
    \text{Protocol 1} :&& \min_{V} \sum_{\a} |(Q_{\a})_{max} - (Q_{\a})_{min}|  \label{eq:opt1}
    \eea
where $Q_{\vec{a}}$ is in the decomposition in Eq. (\ref{eq:dcomq}), and 
\bea
\text{Protocol 2} :&& \max_{V} \sum_{\a} \bra{\a}V^\dagger\ket{\alpha_{1}^{[\vec{a}]}  }.  \label{eq:opt2}
\eea 
Note that local unitaries optimized above according to Protocols 1 and 2 do not coincide in general. 

\begin{figure*}[t]
    \centering
    \includegraphics[width=\textwidth]{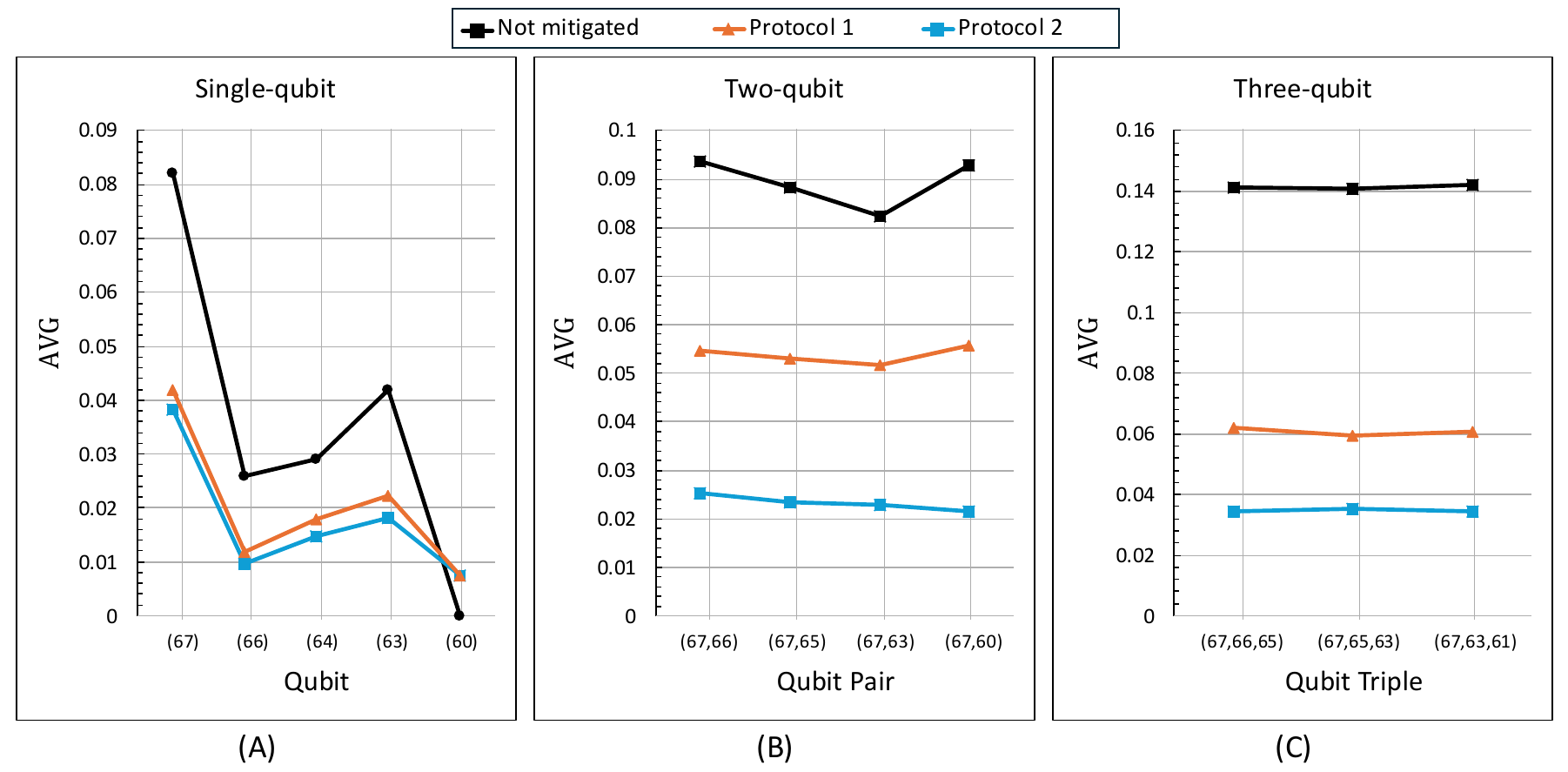}
    \caption{ The average gap between probabilities of noiseless and noisy measurements, denoted by $\mathrm{AVG}$, is given by $\sum_{\vec{a}\in A} |p_0(\vec{a}) - p (\vec{a})|/|A|$ where $|A|$ is the cardinality. The gap is close to $10\%$ without error mitigation (black). Protocol 1 (orange) and Protocol 2 (blue) significantly suppress measurement error rates. (A) Protocols 1 and 2 are applied to measurement errors of individual qubits, labeled 60, 63, 64, 66, and 67. Qubit labeled 60 has almost no measurement error, for which mitigation protocols may give an effect of noise. We recall that the proposed mitigation protocols suppress error rates up to $10^{-2}$. (B) Crosstalk noise may be present, and two protocols significantly reduce error rates. (C) Measurement error rates across three qubits exceed $10\%$, which the proposed mitigation protocols suppress to a percent level. 
}
    \label{fig:compa}
\end{figure*}

Fig. \ref{fig:compa} shows the results of measurement error mitigations where the QPP and the CPP suppress measurement error rates on average. In all cases of single-, two-, and three-qubit measurements, the suppression of measurement errors as well as crosstalk noise is demonstrated.  


%

\section{Conclusion}
\label{sec:c}

In conclusion, we have established the framework for mitigating measurement errors on multiple qubits. The framework consists of two steps: the QPP, which applies a local unitary before noisy measurements, and the CPP, which fixes the probabilities resulting from measurements. We have presented Protocols 1 and 2 by exploiting two decompositions of POVM elements of noisy measurements, one PBD and the other ED. We show that both protocols can suppress crosstalk errors up to a percentage level. As proof-of-principle demonstrations, we have shown measurement error mitigation for two and three qubits in a realistic device with a fake backend `FakeBrisbane' that fully mimics real QPU `ibm\_brisbane'. The mitigation protocols are universal in that they apply to measurements across multiple qubits by suppressing all types of measurement errors, such as single-qubit individual and crosstalk noise. 

The proposed protocols are practical and useful. All the resources required to implement the mitigation protocol, e.g., local unitary operations, are feasible with currently available quantum technologies. Our results pave the way for suppressing measurement errors that persist in present-day quantum hardware. Importantly, measurement error mitigation can enhance quantum algorithms that repeatedly apply state preparation and measurements, such as variational quantum algorithms, including quantum approximate optimization algorithms \cite{bharti2021noisy}, which are well-suited to noisy quantum hardware; it can also prevent measurement errors from propagating through multiple layers. 

For future directions, it would be interesting to investigate the correlations in noisy quantum measurements and their relationship to the effect of measurement error mitigation across multiple qubits. In addition, as shown in Ref. \cite{9142431}, it would be desirable to circumvent QDT to realize measurement error mitigation on multiple qubits, since the computational cost of QDT for multiple qubits is demanding due to its complexity. 


\section*{Acknowledgements}
This work is supported by the National Research Foundation of Korea (RS-2024-00408613) and the Institute for Information \& Communication Technology Promotion (IITP) (RS-2023-00229524, RS-2025-02304540, RS-2025-25464876, RS-2025-25464616). H.-Y. K. is supported by the National Science and Technology Council (NSTC), (with grant number NSTC 112-2112-M-003-020-MY3).

\bibliographystyle{IEEEtran} 
\bibliography{refs}

@article{PhysRevLett.127.090502,
	author = {Geller, Michael R.},
	doi = {10.1103/PhysRevLett.127.090502},
	journal = {Phys. Rev. Lett.},
	numpages = {6},
	pages = {090502},
	publisher = {American Physical Society},
	title = {Conditionally Rigorous Mitigation of Multiqubit Measurement Errors},
	volume = {127},
	year = {2021}}

@article{Geller_2020,
	author = {Geller, Michael R},
	doi = {10.1088/2058-9565/ab9591},
	journal = {Quantum Sci. Technol.},
	pages = {03LT01},
	publisher = {IOP Publishing},
	title = {Rigorous measurement error correction},
	volume = {5},
	year = {2020}}

@article{Wang:2023aa,
	author = {Wang, Kun and Chen, Yu-Ao and Wang, Xin},
	doi = {10.1007/s11432-023-3786-1},
	id = {Wang2023},
	journal = {Sci. China Inf. Sci.},
	number = {8},
	pages = {180508},
	title = {Mitigating quantum errors via truncated Neumann series},
	volume = {66},
	year = {2023}}

@article{9645257,
	author = {Seo, Seungchan and Bae, Joonwoo},
	doi = {10.1109/MIC.2021.3133437},
	journal = {IEEE Internet Computing},
	number = {1},
	pages = {26-33},
	title = {Measurement Crosstalk Errors in Cloud-Based Quantum Computing},
	volume = {26},
	year = {2022}}

@article{PRXQuantum.2.040338,
	author = {Rudinger, Kenneth and Hogle, Craig W. and Naik, Ravi K. and Hashim, Akel and Lobser, Daniel and Santiago, David I. and Grace, Matthew D. and Nielsen, Erik and Proctor, Timothy and Seritan, Stefan and Clark, Susan M. and Blume-Kohout, Robin and Siddiqi, Irfan and Young, Kevin C.},
	doi = {10.1103/PRXQuantum.2.040338},
	journal = {PRX Quantum},
	numpages = {21},
	pages = {040338},
	publisher = {American Physical Society},
	title = {Experimental Characterization of Crosstalk Errors with Simultaneous Gate Set Tomography},
	volume = {2},
	year = {2021}}

@article{PhysRevApplied.17.014024,
	author = {Lienhard, Benjamin and Veps\"al\"ainen, Antti and Govia, Luke C.G. and Hoffer, Cole R. and Qiu, Jack Y. and Rist\`e, Diego and Ware, Matthew and Kim, David and Winik, Roni and Melville, Alexander and Niedzielski, Bethany and Yoder, Jonilyn and Ribeill, Guilhem J. and Ohki, Thomas A. and Krovi, Hari K. and Orlando, Terry P. and Gustavsson, Simon and Oliver, William D.},
	doi = {10.1103/PhysRevApplied.17.014024},
	journal = {Phys. Rev. Appl.},
	numpages = {18},
	pages = {014024},
	publisher = {American Physical Society},
	title = {Deep-Neural-Network Discrimination of Multiplexed Superconducting-Qubit States},
	volume = {17},
	year = {2022}}

@article{TILLY20221,
	author = {Jules Tilly and Hongxiang Chen and Shuxiang Cao and Dario Picozzi and Kanav Setia and Ying Li and Edward Grant and Leonard Wossnig and Ivan Rungger and George H. Booth and Jonathan Tennyson},
	doi = {https://doi.org/10.1016/j.physrep.2022.08.003},
	journal = {Physics Reports},
	pages = {1-128},
	title = {The Variational Quantum Eigensolver: A review of methods and best practices},
	volume = {986},
	year = {2022}}

@article{BLEKOS20241,
	author = {Kostas Blekos and Dean Brand and Andrea Ceschini and Chiao-Hui Chou and Rui-Hao Li and Komal Pandya and Alessandro Summer},
	doi = {https://doi.org/10.1016/j.physrep.2024.03.002},
	journal = {Physics Reports},
	pages = {1-66},
	title = {A review on Quantum Approximate Optimization Algorithm and its variants},
	volume = {1068},
	year = {2024}}

@article{bharti2021noisy,
	author = {Kishor Bharti and Alba Cervera-Lierta and Thi Ha Kyaw and Tobias Haug and Sumner Alperin-Lea and Abhinav Anand and Matthias Degroote and Hermanni Heimonen and Jakob S. Kottmann and Tim Menke and Wai-Keong Mok and Sukin Sim and Leong-Chuan Kwek and Al{\'a}n Aspuru-Guzik},
    doi = {10.1103/RevModPhys.94.015004},
    publisher = {American Physical Society},
	journal = {Rev. Mod. Phys.},
	pages = {015004},
	title = {Noisy intermediate-scale quantum algorithms},
	volume = {94},
    issue = {1},
	year = {2022}}

@article{ibmq,
	author = {See https://quantum-computing.ibm.com/}}

@article{PhysRevA.100.052315,
	author = {Chen, Yanzhu and Farahzad, Maziar and Yoo, Shinjae and Wei, Tzu-Chieh},
	doi = {10.1103/PhysRevA.100.052315},
	journal = {Phys. Rev. A},
	numpages = {17},
	pages = {052315},
	publisher = {American Physical Society},
	title = {Detector tomography on IBM quantum computers and mitigation of an imperfect measurement},
	volume = {100},
	year = {2019}}

@article{9142431,
  author={Kwon, Hyeokjea and Bae, Joonwoo},
  journal={IEEE Trans. Comput.}, 
  title={A Hybrid Quantum-Classical Approach to Mitigating Measurement Errors in Quantum Algorithms}, 
  year={2021},
  volume={70},
  number={9},
  pages={1401-1411},
  doi={10.1109/TC.2020.3009664}}

@article{Maciejewski2020mitigationofreadout,
	author = {Maciejewski, Filip B. and Zimbor{\'{a}}s, Zolt{\'{a}}n and Oszmaniec, Micha{\l{}}},
	journal = {Quantum},
	pages = {257},
	title = {Mitigation of readout noise in near-term quantum devices by classical post-processing based on detector tomography},
	volume = {4},
	year = {2020}}

@article{Sarovar2020detectingcrosstalk,
	author = {Sarovar, Mohan and Proctor, Timothy and Rudinger, Kenneth and Young, Kevin and Nielsen, Erik and Blume-Kohout, Robin},
	journal = {{Quantum}},
	pages = {321},
	title = {Detecting crosstalk errors in quantum information processors},
	volume = {4},
	year = {2020}}

@article{gottesman,
	author = {Gottesman, Daniel},
	journal = {arXiv:quant-ph/9705052},
	title = {Stabilizer Codes and Quantum Error Correction},
	year = {1997}}

@article{PhysRevA.52.R2493,
	author = {Shor, Peter W.},
	doi = {10.1103/PhysRevA.52.R2493},
	journal = {Phys. Rev. A},
	pages = {R2493--R2496},
	publisher = {American Physical Society},
	title = {Scheme for reducing decoherence in quantum computer memory},
	volume = {52},
	year = {1995}}

@article{kim2025protocolpurifyingnoisypreparation,
      title={Protocol for Purifying Noisy Preparation and Measurements of Qubits}, 
      author={Jaemin Kim and Seungchan Seo and Jiyoung Yun and Benjamin Lienhard and Joonwoo Bae},
      year={2025},
      journal = {arXiv:2508.16136},
}




\begin{IEEEbiography}[{\includegraphics
[width=1in,height=1.25in,clip,
keepaspectratio]{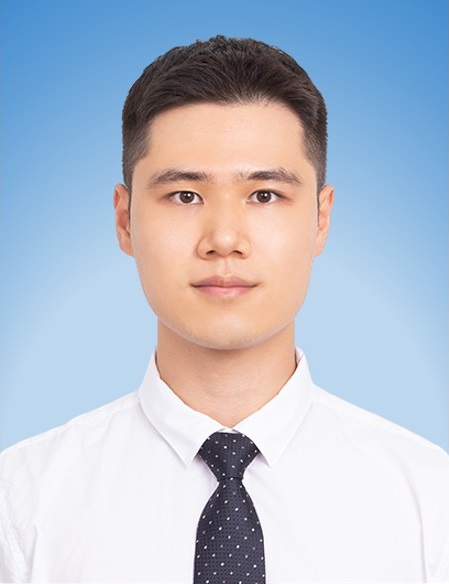}}]
{Seungchan Seo} received the B.E. degree in 2020 from Hanyang University, Seoul, Republic of Korea, and the M.S. degree in 2022 from the Korea Advanced Institute of Science and Technology (KAIST), Daejeon, Republic of Korea, where he is currently pursuing the Ph.D. degree.
His research interests include quantum computing, quantum tomography, quantum benchmarking, and verification of quantum devices.
\end{IEEEbiography}

\begin{IEEEbiography}[{\includegraphics
[width=1in,height=1.25in,clip,
keepaspectratio]{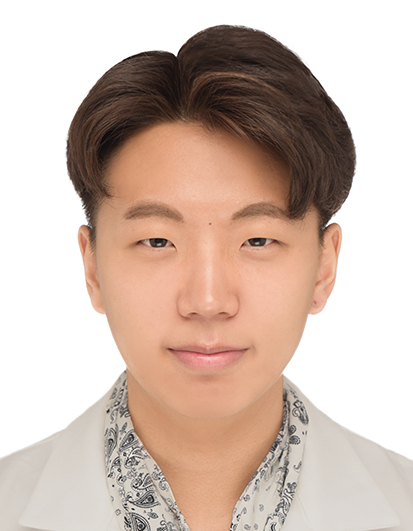}}]{Jiheon Seong}
 received the B.E. degree in 2018 and the M.S. degree in 2020 from the Korea Advanced Institute of Science and Technology (KAIST), Daejeon, Republic of Korea, where he is currently pursuing the Ph.D. degree.
His research interests include entanglement theory, entanglement measures, quantum resource theories, and the implementation and verification of entanglement on quantum hardware platforms.
\end{IEEEbiography}

\begin{IEEEbiography}[{\includegraphics
[width=1in,height=1.25in,clip,
keepaspectratio]{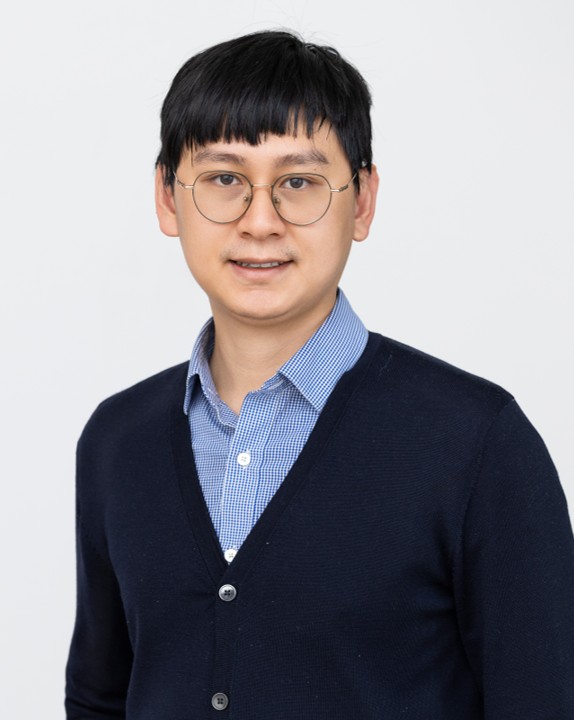}}]{Huan-Yu Ku} received the B.Sc., M.Sc. and Ph.D. degrees in physics from National Cheng-Kung University, Tainan, Taiwan, in 2013 ,2015, and 2019, respectively. From 2014 to 2015, he was a visiting student at Riken, Japan. He has been a Post-Doctoral Research Fellow from 2019 to 2022 at National Cheng-Kung University and from 2022 to 2023 at The Institute for Quantum Optics and Quantum Information (IQOQI), Vienna, Austria. He is currently an Assistant Professor of physics National Taiwan Normal University. His research interests include quantum information, quantum foundations, optimization problems, and quantum resource distillation.
\end{IEEEbiography}

\begin{IEEEbiography}[{\includegraphics
[width=1in,height=1.25in,clip,
keepaspectratio]{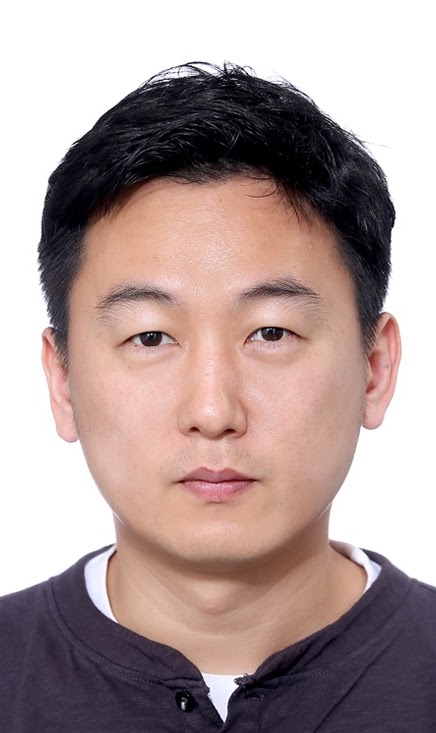}}]{Joonwoo Bae} (Member, IEEE) obtained a Ph.D. in Theoretical Physics from Universitat de Barcelona \& ICFO-Institute of Photonic Sciences, Barcelona in 2007. He has worked at the Korea Institute for Advanced Study (KIAS), Centre for
Quantum Technologies (CQT) in Singapore, the ICFO, Freiburg Institute for Advanced Studies (FRIAS) as a Junior Fellow, and Hanyang University. He is currently with the School of Electrical Engineering, Korea Advanced Institute of Science and Technology (KAIST). His research interests include secure quantum communication, entanglement applications, open quantum systems, quantum foundations, and their practical applications.
\end{IEEEbiography}

\end{document}